\documentclass{article}[16pt]
\usepackage[utf8]{inputenc}
\usepackage[english]{babel}
\usepackage{graphicx}
\usepackage{textcomp}
\begin{document}

\begin{center}
{\Large \bf  Search for dwarf galaxies in the southwestern sector of the Local cosmic void}\\
\bigskip

{\large A.A.Popova$^1$, I.D.Karachentsev$^2$}\\

$^1$Peter the Great St.Petersburg Polytechnic University, Saint Petersburg, 195251 Russia\\
$^2$Special Astrophysical Observatory of the Russian Academy of Sciences, N.Arkhyz, 369167, Russia\\
\end{center}

{\bf Abstract}

  We performed a search for new dwarf galaxies in a direction towards the 
southwestern part of the Local Void using the data on DESI Legacy Imaging
Surveys. In a sky area of $\sim1000$ square degrees, we discovered 12 candidates
to nearby dwarfs with a high confidence. Four of them are probable new 
companions to the nearby galaxy M\,83 and others are isolated objects. We 
found also 20 nearby dwarf candidates with a low confidence. Almost all of
the detected galaxies are classified as late type dwarfs. A new cluster of bluish
stars with an angular diameter of 0.9$^{\prime}$ is revealed by us at a  high galactic 
latitude, $b = -29^{\circ}$. Being at a distance of $\sim70$ kpc, it can be a globular cluster 
associated with the Milky Way stellar stream Sagittarius dSph or a new ultra-faint satellite of the Milky Way.

\bigskip


\bigskip

\section{Introduction} Cosmic voids, along with cosmic filaments, walls, and clusters, are one of the 
fundamental elements of the large-scale structure of the universe. The nearest Local Void, discovered 
by Tully (1988), extends from the boundary of the Local Group of galaxies ($\sim1$~Mpc) to a distance 
of $D\sim20$~Mpc and occupies a sky region with a diameter of about 80$^{\circ}$. According to observational 
data (Karachentseva et al. 1999, Kraan-Korteweg et al. 2008, Tikhonov \& Karachentsev 2006), the geometric 
center of the Local Void has equatorial coordinates near RA = 19.0$^h$, Dec. =+3$^{\circ}$, in a 
region of strong interstellar extinction.
Surveys of this area in the 21 cm neutral hydrogen line and infrared range (Zwaan et al. 2005, 
Kraan-Korteweg et al. 2008, Giovanelli et al. 2005, Roman et al. 1996, Nakanishi et al. 1997) have 
shown that the low number of galaxies in the Local Void zone with radial velocities $V_h<1500$~km s$^{-1}$ is 
real and not solely due to light absorption in the Milky Way.

According to Nasonova \& Karachentsev (2011), the mean number density of galaxies in the Local Void is five 
times lower than the global mean density. Among the sparse population of the Void, only gas-rich dwarf 
galaxies with active star formation (Kreckel et al. 2011) are observed, so the mean stellar mass density 
in the Void is more than an order of magnitude lower than the global mean stellar density.

Searches for galaxies in the Local Void region using photographic plates from the Palomar Sky Survey 
(Karachentseva et al. 1999) and optical identifications of 21-cm line radio sources (Donley 
et al. 2005, Schr\"{o}der et al. 2019) led to the discovery of several nearby dwarf galaxies in the 
heart of the Local Void: KK\,246, HIZOA\,J1914+10, EZOA\,J2129+52 at distances within 10~Mpc. With the 
publication of data from the deep DESI Legacy Imaging Surveys (Dey et al. 2019), it became possible 
to search for new dwarf galaxies in the Local Void. Using this data in a northern Local Void region with 
coordinates RA $> 13.5^h$, Dec. = [$-5^{\circ},+60^{\circ}$], bounded on the east by the Milky Way 
belt, Karachentsev et al. (2024) identified a dozen candidates for membership in the Void. 
Subsequent radial velocity measurements of these galaxies (Karachentsev et al. 2025, Nazarova et 
al. 2025) revealed that six of them are very nearby objects with heliocentric velocities $V_h<200$~km s$^{-1}$.

The aim of our work is to continue the search for new nearby dwarf galaxies in the southwestern 
part of the Local Void using data from the DESI survey (Dey et al. 2019). In the future, we plan to extend the 
search to the eastern side of the Void.

\section{Nearby galaxies in the SW-sector of the Local Void} The map in Fig. 1 presents the sky distribution 
in equatorial coordinates of 5000 nearby galaxies with radial velocities $V_h<1500$~km s$^{-1}$. Galaxies 
of different morphological types are marked with colors corresponding to de Vaucouleurs digital classification shown
in the upper left side of the figure. Objects with active and quenched star formation are labeled blue 
and red, respectively. A circle near the center of the figure outlines the virial zone of the Virgo Cluster. 
The dashed line marks the northwestern region of the Local Void already surveyed by DESI (Karachentsev et al. 2024). 
The solid line depicts the boundary of the area where we conducted our search for new dwarfs.

\begin{figure}[htbp]
\center{\includegraphics[width=1\textwidth]{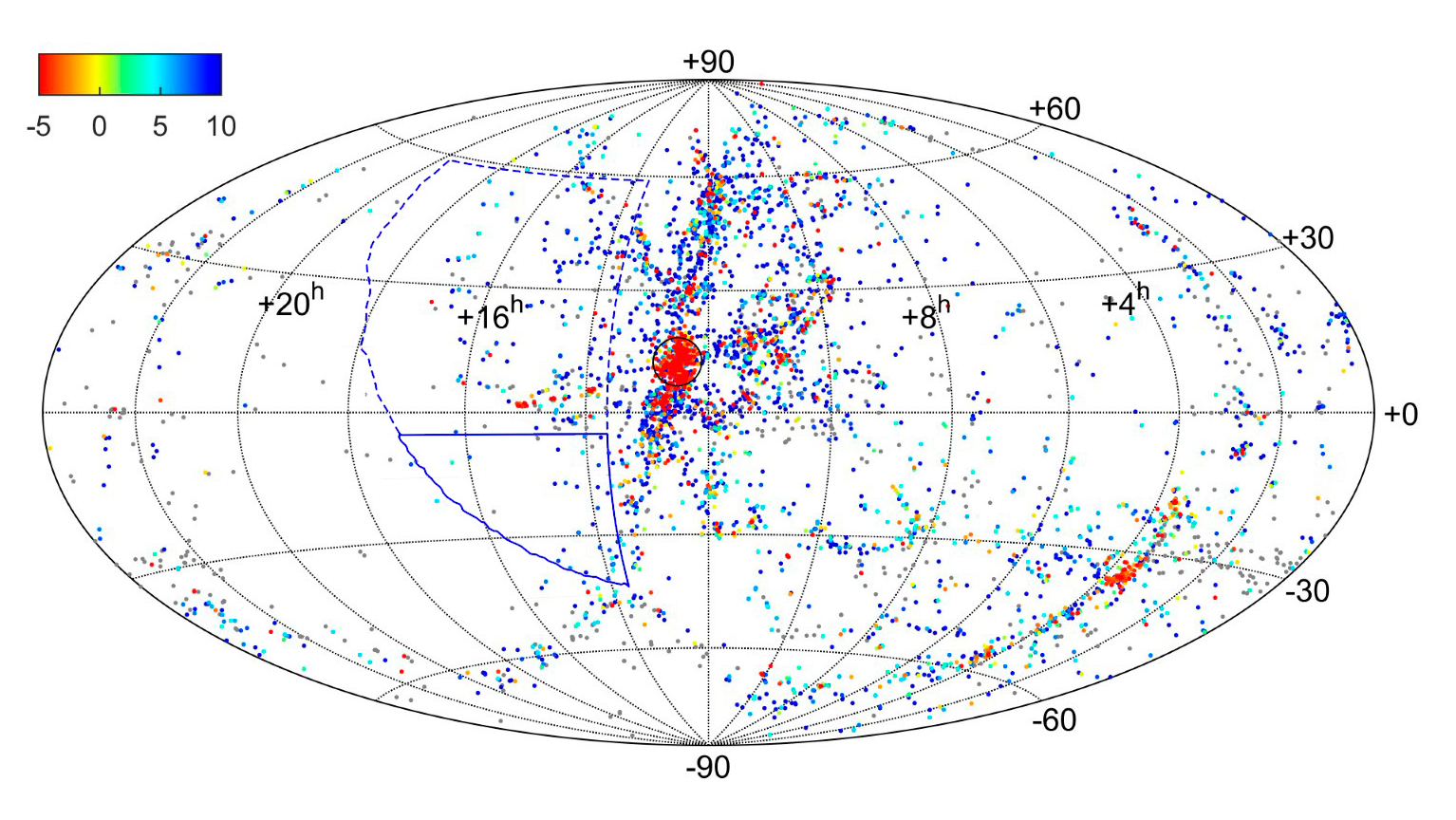}}
\caption{The distribution of galaxies in the Local Supercluster with heliocentric radial velocities $V_h<1500$~km s$^{-1}$ in equatorial coordinates. Early-type galaxies (red) and late-type galaxies (blue) are color-coded according to de Vaucouleurs morphological classification system. The circle at the center marks the virial radius of the Virgo Cluster. The dashed line outlines the region of the Local Void previously surveyed. The solid line denotes the area of the current survey.}
\label{fig:figure1}
\end{figure}

Table 1 lists 28 galaxies in the Local Volume with distances less than $D\sim12$~Mpc, located in the studied 
SW sector of the Local Void. The table columns include: (1)~--- the galaxy name as designated in the Updated Nearby 
Galaxy Catalog (UNGC, Karachentsev et al. 2013); (2)~--- equatorial coordinates of the object in degrees for 
epoch J2000.0; (3)~--- galaxy distance in Mpc; (4)~--- method used to estimate the distance (``TRGB''~--- tip 
of the red giant branch, ``TF''~--- Tully-Fisher relation between the 21-cm line width and galaxy luminosity 
(Tully et al. 2008), ``mem''~--- probable membership in a galaxy group).

\begin{table}[ht]
\centering
\caption{LV galaxies in the Local Void  SW area} 
\begin{tabular}{lcrc} \hline

 Name         &     RA (2000.0)  Dec &   Dist.  &  method \\ \hline
(1)           &            (2)       &    (3)   &   (4) \\ \hline
 IC 4316      &      205.075-28.894  &    4.35  &    TRGB \\
 dw1340-30    &      205.079-30.359  &    5.06  &    TRGB \\
 dw1341-33    &      205.304-33.825  &    4.9   &     mem \\
 NGC5264      &      205.404-29.913  &    4.79  &    TRGB \\
 dw1341-43    &      205.404-43.854  &    3.53  &    TRGB \\
 $[$KK2000$]$ 57  &      205.408-42.581  &    3.84  &    TRGB \\
 dw1342-43    &      205.683-43.255  &    2.9   &     TRGB\\
 KK213        &      205.899-43.769  &    3.77  &    TRGB \\
 ESO 383-070  &      206.230-35.413  &   10.2   &    TF   \\
 ESO325-011   &      206.253-41.858  &    3.4   &     TRGB\\
 $[$KK2000$]$ 58  &      206.503-36.328  &    3.75  &    TRGB \\
 KK218        &      206.664-29.979  &    4.94  &    TRGB \\
 HIPASS1348-37&      207.141-37.967  &    5.65  &    TRGB \\
 ESO383-087   &      207.328-36.061  &    3.19  &    TRGB \\
 ESO 383-091  &      207.632-37.289  &   11.6   &    TF   \\
 PGC 623912   &      207.808-37.391  &   11.6   &    mem  \\
 dw1357-28    &      209.250-28.920  &    4.9   &     mem \\
 ESO384-016   &      209.256-35.333  &    4.49  &    TRGB \\
 NGC5398      &      210.342-33.064  &   12.42  &    TRGB \\
 dw1401-32    &      210.354-32.629  &    4.9   &      mem\\
 dw1403-33    &      210.825-33.403  &    4.9   &      mem\\
 NGC5408      &      210.839-41.376  &    5.32  &    TRGB \\
 dw1406-29    &      211.670-29.136  &    4.9   &     mem \\
 dw1409-33    &      212.262-33.827  &    4.9   &     mem \\
 dw1410-34    &      212.695-34.868  &    4.9   &     mem \\
 dw1413-34    &      213.283-34.392  &    4.9   &     mem \\
 dw1415-32    &      213.920-32.572  &    4.9   &     mem \\
 ESO272-025   &      220.856-44.705  &    3.91  &     TRGB\\
\hline
\end{tabular}
\end{table}

As a supplement, Table 2 provides a list of 27 galaxies in this region from the HyperLEDA database (Makarov et 
al. 2014) with radial velocities $V_h<1500$~km s$^{-1}$. The table columns specify:
(1)~--- galaxy name;
(2)~--- equatorial coordinates in degrees;
(3)~--- radial velocity in km s$^{-1}$;
(4)~--- kinematic distance in the Numerical Action Method (=NAM, Kourkchi et al. 2020) model accounting for 
local galaxy flows (in Mpc);
(5)~--- galaxy distance in Mpc, determined by us via the Tully-Fisher relation (Tully et al. 2008).

\begin{table}[htbp]
\centering
\caption{Galaxies from LEDA with $V_h < 1500$~km~s$^{-1}$}
\begin{tabular}{lcrrr} \hline

    Name     &  RA (2000.0) Dec &  $V_h$  &$D_{\rm NAM}$ & $D_{\rm TF}$ \\ \hline    
  NGC5247    &  204.512-17.884  & 1357 & 16.5 & 11.1 \\
  DDO 180    &  204.543-09.801  & 1299 & 15.8 & 11.2 \\
  ESO577-027 &  205.695-19.581  & 1412 & 17.3 & 19.4 \\
  PGC048786  &  206.408-05.984  & 1453 & 18.8 & 15.4 \\
  PGC947953  &  207.289-12.761  & 1394 & 17.7 & 19.4 \\   
  ESO383-092 &  207.674-35.915  & 1410 & 15.9 &  9.0 \\
  ESO384-002 &  207.828-33.807  & 1390 & 15.9 & 13.0 \\         
  ESO510-015 &  208.763-23.214  & 1357 & 16.4 & 21.4 \\ 
  PGC537199  &  209.416-44.446  & 1475 & 16.7 & 15.5 \\      
  ESO510-052 &  210.894-27.279  & 1327 & 15.8 & 20.1 \\ 
  NGC5510    &  213.405-17.983  & 1438 & 17.5 & 13.5 \\
  NGC5530    &  214.613-43.389  & 1193 & 12.4 & 12.4 \\ 
  NGC5556    &  215.142-29.241  & 1383 & 16.5 & 18.4 \\
  PGC731537  &  215.293-29.251  & 1262 & 13.9 &   -  \\
  ESO446-053 &  215.321-29.263  & 1381 & 16.4 & 13.7 \\    
  PGC890565  &  216.129-16.988  & 1487 & 17.7 & 15.7 \\
  PGC600015  &  216.454-39.267  & 1178 & 12.2 & 12.1 \\ 
  NGC5643    &  218.169-44.174  & 1199 & 12.4 & 13.2 \\
  PGC538542  &  218.220-44.321  & 1078 & 10.7 &   -  \\
  ESO326-029 &  218.276-41.759  & 1204 & 12.5 & 15.4 \\
  ESO327-014 &  220.728-38.014  & 1236 & 13.0 & 17.4 \\ 
  ESO386-013 &  220.766-33.491  & 1376 & 14.8 & 19.9 \\                 
  ESO328-043 &  229.823-41.233  & 1337 & 14.4 & 27.2 \\ 
  ESO274-016 &  232.230-42.783  & 1331 & 14.4 & 27.4 \\            
  PGC3994656 &  239.583-10.535  &  935 & 11.5 & 14.4 \\ 
  PGC057723  &  244.315-11.731  &  977 & 12.1 & 13.9 \\
  PGC165693  &  250.761-20.668  & 1206 & 14.4 & 15.6 \\  
\hline                                               
 \end{tabular}                                       
 \end{table}

Independent distance estimates yield a mean difference of $\langle D_{\rm TF} - D_{\rm NAM}\rangle =+1.14\pm0.90$~Mpc 
and a root mean square difference of $\Delta D=4.67$~Mpc or 32\% at $\langle D_{\rm NAM}\rangle =14.5$~Mpc. For 
a sample of low-luminosity galaxies, this level of distance determination error appears acceptable.
The distribution of galaxies in the studied region is shown in Fig. 2. Local Volume members are marked with 
crosses, while galaxies from LEDA with $V_h<1500$~km s$^{-1}$ are indicated by empty squares. The area not 
covered by the DESI Legacy Imaging Surveys is shaded gray.

Near the lower-right boundary of this region lie two nearby groups around the galaxies M83 (NGC\,5236) and 
Centaurus\,A (NGC\,5128), with central coordinates [204.250, --29.868] and [201.370, --43.017], respectively. 
The M83 galaxy group at a distance of 4.90~Mpc (TRGB) has a virial mass of $M_T=1.07\times 10^{12} M_{\odot}$ 
(Karachentsev \& Kashibadze 2021). According to the relation from Tully (2015), 
$R_v/215$~kpc $=(M_T/10^{12} M_{\odot})^{1/3}$, the virial radius of the group, $R_v$, is 220~kpc or 
$2.6^{\circ}$, while the zero-velocity surface radius, $R_0\simeq3.5R_v$, beyond which field galaxies 
participate in the general Hubble expansion, spans $9.0^{\circ}$. Both radii are shown in the figure as 
solid and dashed circles. The galaxy group around Cent\,A has a distance of 3.68~Mpc (TRGB), a virial mass 
of $M_T=4.67\times 10^{12} M_{\odot}$ (Karachentsev \& Kashibadze 2021), and radii $R_v=5.6^{\circ}, 
R_0 =19.6^{\circ}$.

\begin{figure}[htbp]
\includegraphics[width=1\textwidth]{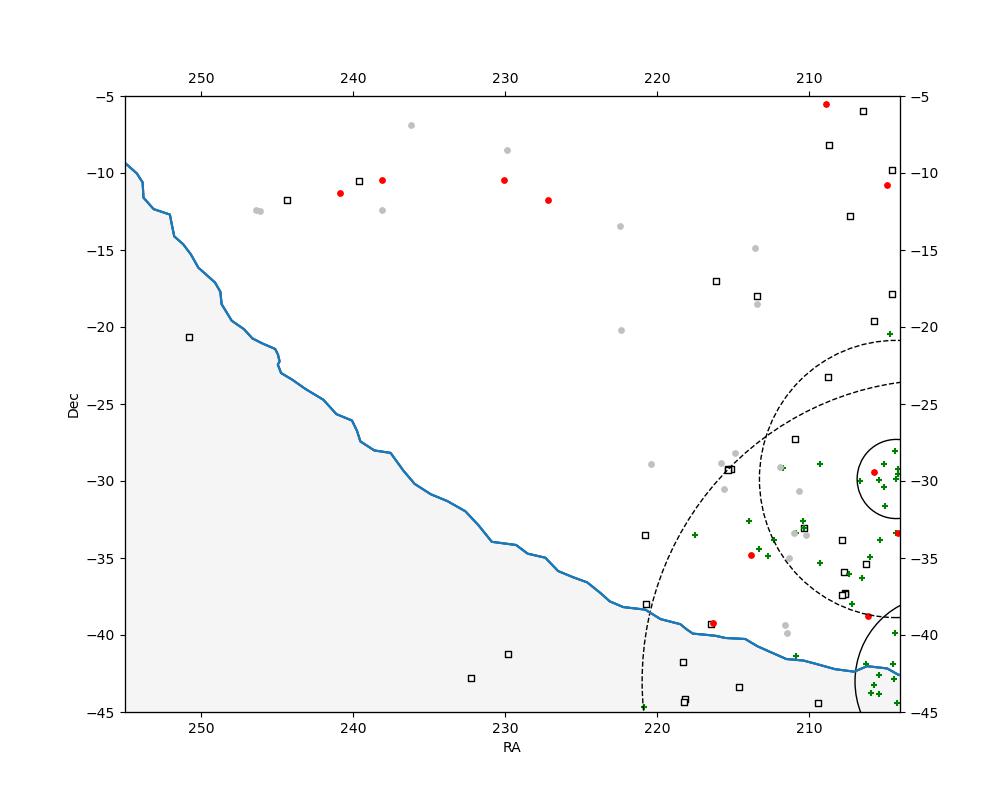}
\caption{Our field of view is in equatorial coordinates. The gray region marks the zone outside the Legacy Surveys. Solid and dashed circles correspond to the virial radius and zero-velocity radius of the galaxy groups around M83 and Centaurus A. LV members are shown as crosses, galaxies from LEDA with $V_h<1500$~km s$^{-1}$ are marked with empty squares.New dwarf galaxies detected with high and low confidence are indicated by red and gray circles, respectively.}
\label{fig:figure2}
\end{figure}
 
 In addition to members of these two nearby groups, at the far boundary of the Local Volume lies the 
low-luminosity spiral galaxy NGC\,5398, located at a distance of 12.42~Mpc (TRGB, Tully et al. 2009), and 
three galaxies: ESO\,383-070, ESO\,383-091, PGC\,623912, with less precise distance estimates.

The portion of the region we studied with approximate coordinates RA $<220^{\circ}$ and Dec. $<-20^{\circ}$ was thoroughly investigated by M\"{u}ller et al. (2015, 2017), who discovered numerous dwarf satellites of 
M83 and Cent\,A there.
  
  \section{Searches for new nearby dwarfs} By examining galaxy images in the DESI Legacy Imaging Surveys, 
DR10 (Dey et al. 2019) over an area of $\sim$1000 square degrees, we selected galaxies with supposed distances 
less than 15~Mpc. When searching for nearby dwarf galaxies, the focus was on low surface brightness objects 
with signs of structural granularity. Reproductions of 12 dwarf galaxies from the DESI survey are shown in 
Fig. 3. Each image measures $2^{\prime}\times2^{\prime}$, with north at the top and east to the left.

\begin{figure}[htbp]
\includegraphics[width=1\textwidth]{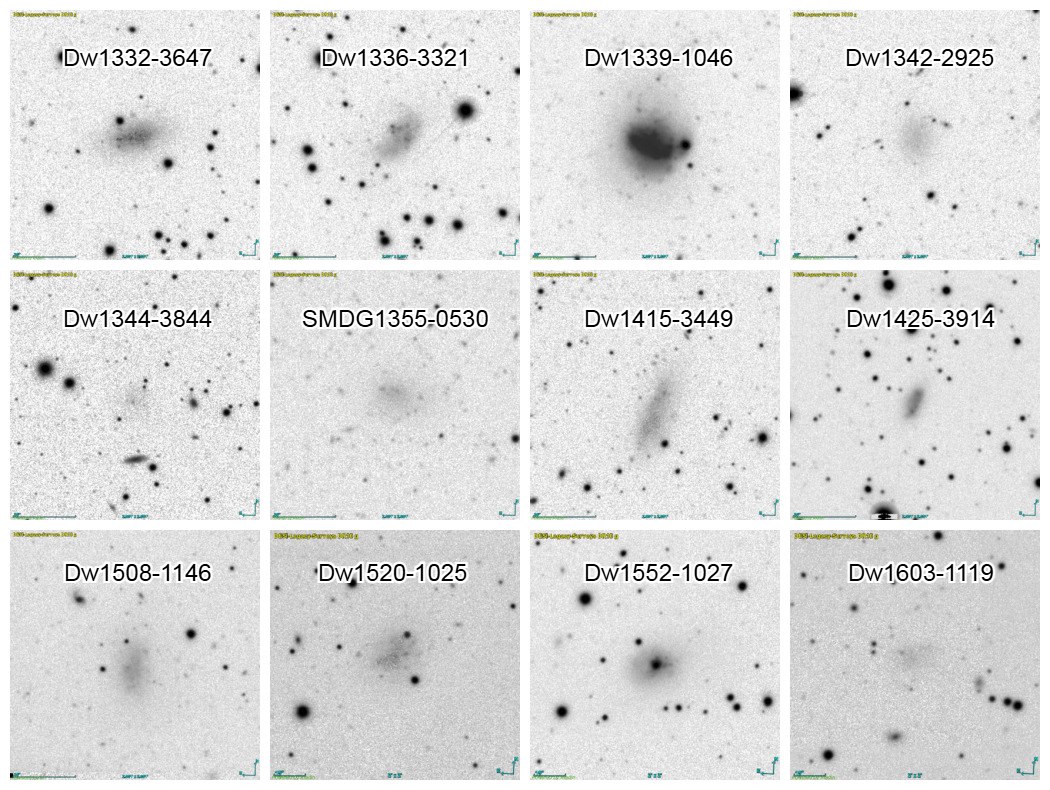}
\caption{Image of 12 new dwarf galaxies from Legacy Surveys found in the Local Void region. Each image size is $2^{\prime}\times2^{\prime}$. North is on the top and East is to the left.}
\label{fig:figure3}
\end{figure}

Data for these galaxies are listed in Table 3, whose columns include:
(1)~--- galaxy name;
(2)~--- equatorial coordinates in degrees;
(3, 4)~--- apparent $g$ and $r$ magnitudes extracted from DESI survey data;
(5)~--- maximum apparent angular diameter of the galaxy in arcminutes.
In two cases, the letter ``i'' marks galaxies lacking $r$-band measurements, with the $i$-band magnitude 
provided instead. Based on their coordinates and image morphology, four galaxies: Dw\,1332-3647, Dw\,1336-3321, 
Dw\,1342-2925 and Dw\,1344-3844 likely new members of the M83 group, while the others are associated with more 
distant background structures.

\begin{table}
\centering
\caption{New Local Void candidates.} 
\begin{tabular}{lcclc} \hline
  Name       & RA (2000.0) Dec   &   $g$   &    $r$ &    $a^{\prime}$ \\ \hline
  Dw1332-3647&    203.152-36.795 &   18.20 &  17.70 &   0.74 \\
 Dw1336-3321 &   204.156-33.357  &  18.42  & 18.10  &  0.53 \\
 Dw1339-1046 &   204.876-10.777  &  16.09  & 15.69i &  1.00 \\
 Dw1342-2925 &   205.713-29.427  &  19.46  & 18.87  &  0.48 \\
 Dw1344-3844 &   206.114-38.734  &  20.63  & 20.11  &  0.40 \\ 
 SMDG1355-05 &   208.898-05.505  &  19.40  & 18.76  &  0.86 \\ 
 Dw1415-3449 &   213.784-34.819  &  19.03  & 18.53  &  0.95 \\
 Dw1425-3914 &   216.279-39.248  &  17.84  & 17.49  &  0.52  \\
 Dw1508-1146 &   227.183-11.777  &  17.94  & 17.38  &  0.75 \\
 Dw1520-1025 &   230.057-10.430  &  17.82  & 17.25  &  0.66 \\
 Dw1552-1027 &   238.072-10.454  &  16.76  & 16.79  &  0.59 \\
 Dw1603-1119 &   240.832-11.327  &  18.92  & 17.89i &  0.41 \\ 
\hline                                                       
 \end{tabular}                                              
 \end{table}
 
In addition to the 12 candidates for new members of the Local Void, we also identified 20 dwarf galaxies 
listed in Table 4. The column designations in this table are the same as in the previous one. The probability 
that these objects reside within the Void is significantly lower than for the dwarf galaxies in Table 3.
  
  According to our classification, nearly all dwarf galaxies listed in Tables 3 and 4 belong to late morphological types: irregular (Irr), Magellanic (Im), and blue compact dwarfs (BCD). Their observed color indices fall within the range $(g-r)$ = [0.03,+0.65], with a median value of +0.40. After correction for interstellar extinction the median color value of +0.33 matches the mean color of \{Irr, Im, BCD\}-dwarfs, which are dominated in regions with extremely low galaxy number density.

The only exception is the spheroidal dwarf galaxy of extremely low surface brightness Dw\,1449-2011 with an 
angular diameter of $0.58^{\prime}$. However, it may be associated with a distant group around S0 galaxies ESO\,580-040 and NGC\,5761, which have radial velocities of $\sim4100$~km s$^{-1}$. If located at the distance 
of this group, $\sim50Mpc$, the spheroidal dwarf would have a linear diameter of 8.4~kpc and would belong to the rare 
category of ultra-diffuse galaxies, UDG.
   
The candidates for membership in the Local Void identified by us with high and low confidence are 
shown in Fig. 2 as red and gray circles, respectively. Approximately half of the new dwarf galaxies lie 
outside the gravitational influence zone of the nearby massive galaxies M83 and Cent\,A. Their distribution 
is highly inhomogeneous, forming a completely empty region at the center of Fig. 2 with an area of about 
300 square degrees. There are also
no HyperLeda galaxies with radial velocities less than 1500~km s$^{-1}$ in this lagoon. Galaxies with 
declinations Dec. $>-20^{\circ}$ may form a filament (arc) extending approximately $30^{\circ}$. To verify 
this hypothesis, radial velocity and distance measurements for these galaxies are clearly required.

 \begin{table}[ht]
\caption{Low confidence candidates to the Local Void.} 
\begin{tabular}{lcclc} \hline
  Name       & RA (2000.0) Dec   &    $g$   &     $r$  &     $a^{\prime}$ \\ \hline
 Dw1400-3330 &   210.164-33.504  &       -  &        - &     0.42 \\ 
 Dw1402-3037 &   210.650-30.621  &    18.63 &    18.19 &     0.40 \\
 Dw1404-3321 &   211.002-33.362  &    20.05 &    19.37 &     0.31 \\
 Dw1405-3459 &   211.329-34.995  &    20.50 &    19.92 &     0.37 \\
 Dw1405-3951 &   211.459-39.866  &       -  &    -     &     0.40 \\
 Dw1406-3922 &   211.550-39.376  &       -  &    -     &     0.41\\
 Dw1407-2905 &   211.916-29.087  &    18.34 &    17.92 &     0.47 \\
 Dw1413-1830 &   213.422-18.502  &    17.96 &    17.56 &     0.53 \\
 Dw1414-1454 &   213.562-14.900  &    17.50 &    17.25 &     0.36  \\ 
 Dw1419-2811 &   214.839-28.188  &       -  &     -    &     0.34  \\   
 Dw1422-3031 &   215.564-30.522  &    19.18 &    18.95 &     0.62  \\
 Dw1423-2850 &   215.767-28.838  &       -  &     -    &     0.50  \\
 Dw1441-2855 &   220.409-28.920  &    19.49 &    19.14 &     0.52  \\
 Dw1449-2011 &   222.365-20.198  &       -  &     -    &     0.58 \\
 Dw1449-1326 &   222.423-13.445  &    18.95 &    18.63 &     0.45  \\ 
 Dw1519-0830 &   229.846-08.511  &    17.74 &    17.36 &     0.56  \\
 Dw1544-0652 &   236.190-06.872  &    17.06 &    16.68 &     0.83  \\
 Dw1552-1222 &   238.091-12.372  &       -  &    -     &     0.34  \\
 Dw1624-1226 &   246.141-12.443  &    17.45 &    16.64i&     0.46  \\
 Dw1625-1224 &   246.402-12.416  &    19.66 &    19.04i&     0.41  \\
  \hline                                                          
   \end{tabular}                                                  
   \end{table}
   
   \section{Bluish star cluster 144305.8--280141} While searching for new dwarf galaxies in the SW-sector 
of the Local Void, we discovered an association of bluish stars at coordinates 
$220.774^{\circ}-28.028^{\circ}$ with an angular diameter of $0.9^{\prime}$. It's reproduction from the 
DESI Legacy Imaging Surveys is shown in Fig. 4, where the image size is $2^{\prime}\times2^{\prime}$, 
north~--- at the top, east~--- to the left. This object is absent from the catalog of 10978 star clusters 
and associations in the Milky Way (Bica et al. 2019). Located at a relatively high galactic latitude, 
$b\simeq29^{\circ}$, this stellar group may be a Milky Way globular cluster. In appearance, the new star 
cluster resembles the bluish globular cluster Arp-Madore\,4 (=AM4), whose angular diameter is 
$2.2^{\prime}$. Carraro (2009) constructed a \{$V vs. B-V$\} diagram for AM4 and determined its 
heliocentric distance as $D=33\pm4$~kpc. Assuming equal linear sizes for both clusters, we derive a rough 
distance estimate of $\sim80$~kpc for the new cluster (named ``Alice'').

We utilized stellar photometry in the $g$ and $i$ filters, available in DESI DR10 (www.legacysurvey.org), 
and determined magnitudes for 40 stars in the new cluster. The color–magnitude diagram for these stars is 
shown in Fig. 5. According to Schlafly \& Finkbeiner, the galactic extinction in this direction is 
$A_g=0.359$~mag, and CE $(g-i)=0.175$. Accounting for extinction reveals that the brightest stars in the 
cluster exhibit a bluish color $\langle g-i\rangle\simeq+0.5$. By comparing the color–magnitude diagrams 
of AM4 and the new cluster, and adjusting for differences in extinction, we estimate a heliocentric distance 
of $D=63\pm10$~kpc for the cluster Alice. Hamren et al. (2013) conducted deeper photometry of stars in the 
globular cluster AM4 using images from the Hubble Space Telescope in the F606W and F814W filters, determining 
a distance of $D_{\rm AM4} = 31.2\pm0.4$~kpc. Comparing the tip of their AM4 diagram with the diagram for 
the new cluster’s stars yields a refined distance estimate of $D_{\rm Alice} =74\pm10$~kpc.

\begin{figure}[htbp]
 \centering
\includegraphics[width=0.7\textwidth]{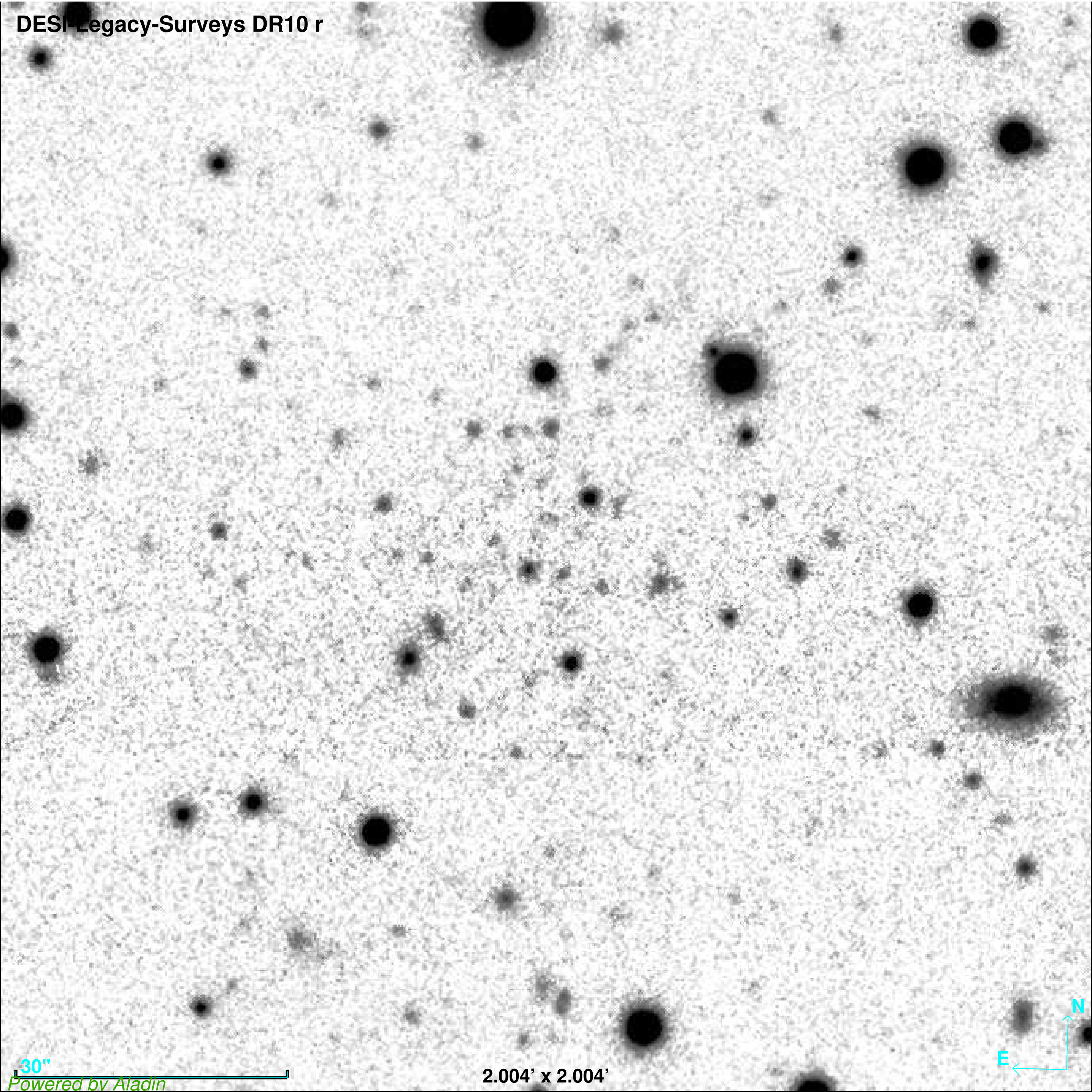}
\caption{Image of the new stellar cluster Alice from Legacy Survey. The image side is $2^{\prime}$. North is on the top and East is to the left.}
\label{fig:figure4}
\end{figure}

Carraro (2009) suggested that the globular cluster AM4 might be part of the Sgr dSph stellar stream, which 
formed due to the tidal disruption of a spheroidal dwarf galaxy. The new globular cluster is located at an 
angular distance of $\sim 11^{\circ}$ from AM4, but its heliocentric distance is more than twice as far. 
Nevertheless, this cluster should also be considered in the context of association with the Sgr dSph stream. Alternatively, it could be a probable dwarf satellite of the Milky Way, similar in distance, galactic latitude and texture to other known ultra-faint satellites, like Kim 2, Eridanus \uppercase\expandafter{\romannumeral3}, Koposov 1 (Pace, 2024). 
To verify whether a connection with the Sph stream exists, a detailed study of the new cluster is required, including 
refining its distance, measuring its radial velocity and proper motion, and determining its chemical 
composition and age.

\begin{figure}[ht]
 \centering
\includegraphics[width=0.425\textwidth]{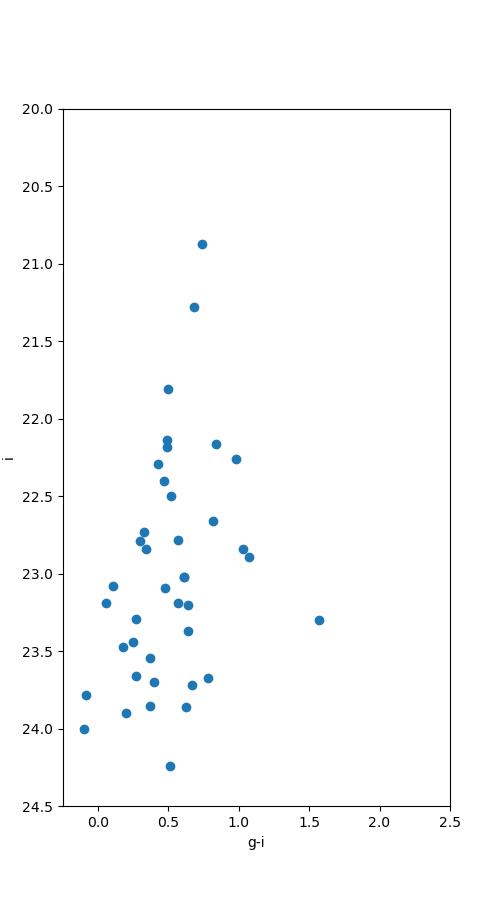}
\caption{CMD of bright stars in the Alice stellar cluster}
\label{fig:figure5}
\end{figure}

   \section{Concluding remark} Our search for new nearby galaxies in the southwestern region of the Local 
Void over an area of approximately 1000 square degrees has led to the discovery of 12 nearby late-type dwarf 
galaxies, including four probable new members of the LV group around M83. In addition, we identified 20 
further candidates for the Void membership with lower confidence in this region. At a high galactic latitude, 
$b=29^{\circ}$, was found a bluish star cluster with an angular diameter of $0.9^{\prime}$. Its preliminary 
distance estimate is $74\pm10$~kpc. This new object is likely a globular cluster that, alongside the known 
globular cluster AM4, may be part of the Sgr dSph stellar stream in the Milky Way or to be an isolated ultra-faint dwarf companion to the Milky Way. Clarifying the nature of 
these objects requires measurements of their radial velocities and distances.

The authors thank A.S. Rastorguev and M.E. Sharina for valuable discussions and advice. This work used data 
from the DESI Legacy Imaging Surveys, DR10, the HyperLEDA database, and the Local Volume Galaxy Database 
(www.sao.ru/lv/lvgdb). The study was supported by the Russian Science Foundation grant № 24--12--00277.

                 {\bf R e f e r e n c e s}

{\em E.Bica, D.B.Pavani, C.J.Bonnato at al.}, Astron. J., {\bf 157}, 12, 2019.
 
{\em G.Carrado},  Astron. J., {\bf 137}, 3809,  2009.  

{\em A.Dey, D.J.Schlegel, D.Lang  et al.}, Astron. J., {\bf 157}, 168 2019.

{\em J.L.Donley,  L.Staveley-Smith, R.C.Kraan-Korteweg et al.}, Astron. J., {\bf 129}, 220, 2005. 

{\em R.Giovanelli, M.P.Haynes,  B.R.Kent et al.}, Astron. J., {\bf 130}, 2598, 2005.

{\em K.M.Hamren, G.H.Smith, P.Guhathakurta  et al.}, Astron. J.,  {\bf 146}, 116, 2013.  

{\em I.D.Karachentsev, M.I.Chazov, S.S.Kaisin}, Mon. Not. Roy. Astron. Soc., {\bf 537L}, 21, 2025.

{\em I.D.Karachentsev, V.E.Karachentseva, S.S.Kaisin, E.I.Kaisina}, Astrophysics, {\bf 66}, 441, 2024.

{\em I.D.Karachentsev, O.G.Kashibadze},  Astron. Nachrichten, {\bf 342}, 999, 2021.

{\em I.D.Karachentsev, D.I.Makarov, E.I.Kaisina}, Astron. J., {\bf 145}, 101, 2013 (UNGC).

{\em V.E.Karachentseva, I.D.Karachentsev, G.M.Richter},  Astron. and Astrophys. Suppl., {\bf 134}, 1, 1999.   

{\em E.Kourkchi, H.M.Courtois, R.Graziani R et al.},  Astron. J., {\bf 159}, 67, 2020.

{\em R.C.Kraan-Korteweg, N.Shafi, B.S.Koribalski  et al.}, in ``Galaxies in the Local Volume'', Astrophysics and Space Science Proceedings, {\bf 5}, 13, 2008.

{\em K.Kreckel, P.J.E.Peebles, J.H.van Gorkom  et al.}, Astron. J., {\bf 141}, 204, 2011.              

{\em D.I.Makarov, P.Prugniel, N.Terekhova  et al.},  Astron. and Astrophys., {\bf 570A}, 13, 2014 (LEDA).

{\em O.M\"{u}ller, H.Jerjen, B.Binggeli},   Astron. and Astrophys., {\bf 597A}, 7, 2017.

{\em O.M\"{u}ller, H.Jerjen, B.Binggeli},   Astron. and Astrophys., {\bf 583A}, 79, 2015.

{\em K.Nakanishi, T.Takata, T.Yamada  et al.}, Astrophys. J. Suppl.,  {\bf 112}, 245, 1997.

{\em O.G.Nasonova \& I.D.Karachentsev},  Astrophysics, {\bf 54}, 1, 2011.

{\em A.E.Nazarova, J.M.Cannon, I.D.Karachentsev  et al.}, AJ, 2025 (in press).

{\em A.B.Pace}, arXiv 2411.07424, 2024.

{\em A.T.Roman, K.Nakanishi, A.Tomita, M.Saito}, Publications of the Astronomical Society of Japan, {\bf 48}, 679, 1996.

{\em E.F.Schlafly, D.P.Finkbeiner}, Astrophys. J., {\bf 737}, 103, 2011.

{\em E.J.Shaya, R.B.Tully, Y.Hoffman, D.Pomarede}, Astrophys. J., {\bf 850}, 207, 2017.

{\em A.C.Schr\"{o}der, L.Fl\"{o}er, B.Winkel, J.Kerp},  Mon. Not. Roy. Astron. Soc., {\bf 489}, 2907, 2019.    

{\em A.V.Tikhonov \& I.D.Karachentsev}, Astrophys. J., {\bf 653}, 969, 2006.

{\em R.B.Tully},  Astron. J., {\bf 149}, 54, 2015.

{\em R.B.Tully, E.J.Shaya, I.D.Karachentsev et al.},  Astrophys. J., {\bf 676}, 184, 2008.   

{\em R.B.Tully, L.Rizzi, E.J.Shaya  et al.}, Astron. J., {\bf 138}, 323, 2009.

{\em R.B.Tully},  Nearby galaxies catalog, Cambridge University Press, Cambridge, 1988.

{\em M.A.Zwaan, L.Staveley-Smith, B.S.Kodibalski  et al.},  Astron. J., {\bf 125}, 2842, 2003.

   \end{document}